\documentstyle[12pt]{article}
\input equal.tex
\begin{document}
\begin{flushright}
GUTPA/91/11-2\\
\end{flushright}
\vskip .2 cm
\begin{center}
{\bf \Large CRITICAL MASS
IN NONZERO TEMPERATURE QCD\\ \vskip .1in
WITH STAGGERED FERMIONS}
\vskip 1.cm
{\Large I.M. Barbour, A. J. Bell and E.G. Klepfish}
\vskip .1in

Dept. of Physics and Astronomy\\
University of Glasgow\\
Glasgow G12 8QQ, U.K.
\end{center}
\vskip .1cm
\baselineskip 28 pt
\begin{center}
\bf Abstract
\end{center}
\noindent
The behaviour of the chiral condensate in QCD is investigated by means
of a study of the distribution of the zeros of the partition function
in the complex quark mass plane. Simulations are performed at fixed
temperature on three different spatial volumes at $\beta=5.04$ and at
$\beta=4.9$ and $\beta=5.2$ on a $4^4$ lattice. Evidence is found for
a chirally related transition at non-zero quark mass in the
intermediate coupling region for $\beta < 5.2 $
but superimposed upon a smooth behaviour
for the condensate. The critical mass at which this transition is
found is only weakly dependent on the spatial volume and decreases
with decreasing temperature.
\newpage
\bigskip
\noindent {\bf 1. Introduction}
\newline\noindent
It is believed that,
at high temperature
and high quark density,
hadronic matter undergoes a phase transition
into a quark-gluon plasma. At the same time it
is expected that the chiral symmetry will be
restored because population of the
states above the Fermi level increases the
energy of the massive states and favours the
chirally symmetric vacuum.
Although this argument was originally
applied by Kogut {\it et al.}
[\ref{KogutMatsuoka}]
to the finite density phase transition in QCD, it also holds
when the higher levels
are populated by increasing the temperature.
\par The renormalization
group analysis
[\ref{PisarskyWilczek}]
implies a first order chiral phase transition
at zero quark mass (for three
or more fermion flavours).
This prediction, based on an effective spin model analysis, also
holds at nonzero quark mass although the phase transition is weakened.
Recent numerical simulations  also show that at
intermediate couplings there exists a "critical mass" at which the
condensate $<\bar\psi\psi>$ fluctuates
between two distinct levels, indicating a possible first
order phase transition
at $m_q\neq 0$
[\ref{Gupta1}],
[\ref{Gupta2}],
[\ref{Gupta3}],
[\ref{Fuku}],
[\ref{KogutSinclair}],
[\ref{Sinclair}],
[\ref{Kogut1}],
[\ref{Bitar}].
At the same time a sharp crossover behaviour
was observed in the deconfinement order parameter.
\par  It is not clear if
these results can be interpreted as
a continuation of the chirally symmetric phase
into a nonzero quark
mass region or as a finite volume effect
masking a true transition at $m_q=0$.
The correspondence between lattice QCD and the continuum theory
of the strong interaction is valid only if such phenomena as
confinement and the spontaneous breaking of chiral symmetry
can be traced in a lattice calculation. Since it is a finite
system, the lattice cannot exhibit any discontinuity of an order
parameter and the appearance of a zero mass Goldstone boson.
However, introducing
a finite mass one can
drive the system into the regime of a broken chiral symmetry
with the corresponding behaviour of physical observables.
In that respect it has been noted by
Jolicoeur and Morel
[\ref{JoliceurMorel}] (in a zero temperature strong coupling study)
that the quantity controlling the chiral
symmetry breaking as well as the expected
functional dependence of the pion mass on the lattice mass, $ma$,
\begin{equation}
m_\pi a \propto \sqrt{ma}+{\cal{O}}({1/L^d})
\end{equation}
is given
by the ratio of two Bessel functions with argument
$s=2ma\times L^d \times \sqrt{2d}\times <\bar\psi\psi>_{\infty}$
with $d$ the space-time dimension, in our case $3+1$.
For fixed volume ($L^d$),
$ma$ has to be big enough such that the $s\rightarrow \infty$
approximation of this ratio, namely,
$1-1/2s+{\cal{O}}(s^{-2})$,
holds. For $ma$ (and thus $s$) much
smaller, the finite volume effects will
alter the results from those expected in the thermodynamical
limit.
Similar conclusions were derived at finite
temperature via a chiral perturbation theory analysis
[\ref{GasserLeutwyller}].
\par Hence the order of taking the limit of infinite volume
and zero mass is crucial.
This is an inherent feature of a finite
system and
the study of the thermodynamics of hadrons from lattice
simulations requires scrupulous analysis of the finite-size
effects, including independent simulations with different
mass parameters and on different lattice sizes.
\par The issue of the mass dependence of any
finite temperature phase transition in lattice
QCD is strongly related to the existence
of a volume independent critical
mass and its dependence on the temperature. There are two
possible scenarios for the phase structure
with massive quarks. There may be no phase
transition for $m_q\neq 0$ or it may remain
but not necessarily with the same strength as at
$m_q=0$.
\par In this paper we determine the critical lattice mass
$m_ca$, at which lattice QCD simulates a phase
transition for several intermediate values of
the inverse gauge coupling constant $\beta$
and, for $\beta=5.04$, analyse its dependence on the lattice
volume.
The method used is similar to that used
by Barbour and Bell
[\ref{BarbourBell}] at strong coupling and is based upon
an analysis of the zeros of the partition
function of lattice QCD in the complex mass plane.
As suggested by Yang and Lee
[\ref{YangLee}], the existence of
a phase transition in the thermodynamical limit is detected by
the behaviour of the zeros of the partition function close to
the positive real axis.
The real part of the zero closest to the real axis (if any exist) is
identified as the critical value of the
lattice mass, but the existence and nature of any
phase transition follows from the behaviour
of the density of zeros in that region as the lattce
spatial volume is increased[\ref{Itzykson}].
\par The chiral condensate is defined as
the derivative of the logarithm
of the partition function $Z(ma)$ and can be expressed
in terms of its zeros, $z_i$ as:
\begin{equation}\label{sumroots}
<\bar\psi\psi>={1\over {n_s}}\sum_i{1\over {ma-z_i}}.
\end{equation}
where $n_s$ is the number of lattice sites. We show below that the
zeros appear as $\pm z_i, \pm z_i^\star$
and
hence[\ref{YangLee}],
[\ref{BarbourDagotto}]
$<\bar\psi\psi(ma)>$ is dual to the
two-dimensional electrostatic field at the point
$ma$ arising from unit positive charges located at $\{z_i\}$.
\par The paper is organized as follows: In section 2 we describe
the method used and present
the numerical results at $\beta=4.9,5.04$ and $5.2$ on a $4^4$
lattice. In Section 3 we study the
behaviour of the distribution of zeros at $\beta=5.04$ with
increasing lattice spatial volume.
\bigskip
\newline
\noindent{\bf 2. The method and results
at intermediate coupling on a $4^4$ lattice}
\bigskip
\newline\noindent
The partition function, $Z(ma)$, for lattice QCD with staggered
fermions is given
[\ref{BarbourBell}],
[\ref{Tassos}]
by:
\begin{equation}\label{Partition}
<{{D(ma)}\over {D(m_0a)}}>
\end{equation}
where the average is taken over an
ensemble of configurations generated at lattice mass
$m_0a$, and $D$ is the determinant of the fermion matrix
\begin{equation}\label{one}
M(ma)=M(0)+Ima
\end{equation}
with $M(0)$ being an
antihermition off-diagonal matrix
connecting nearest neighbours lattice
sites.
Independent field configurations
are generated by means of the Hybrid Monte Carlo algorithm and the
eigenvalues of the matrix $M(0)$ evaluated by using the
Lanczos algorithm. On the larger lattices, $6^3 \times 4$ and
$8^3 \times 4$, we use Lanczos without reorthogonalization[\ref{Lan}]
Checks are performed on the number of converged eigenvalues found,
their degree of convergence ( via their behaviour when the
last row and column of the tridiagonal form are removed),
and using $\sum \lambda_i^2 = 8n_s n_c$, where the $\lambda_i$
are the eigenvalues of $M(0)$, to ensure
that the true set of $\lambda_i$ are found. This method can be
extended to larger lattices.
\par The coefficients of
of the characteristic polynomial of $M$ are simply related to its
eigenvalues and each, normalized with
respect to $D(m_0a)$, is averaged over the ensemble. Since the
$\lambda_i$ are imaginary and have a $\pm$ symmetry, the
coefficients of each characteristic polynomial are always real
and positive
and hence the zeros appear as $\pm z_i, \pm z_i^{\star}$.
This averaged
characteristic polynomial then gives the explicit dependence of the
partition function on $ma$.
In fact it is a polynomial in $(ma)^2$
because of the nearest neighbour structure of the fermion matrix.
The coefficients of this polynomial are $ma$-independent and
should not be affected (apart from an overall normalization factor)
by the choice of the mass parameter $m_0a$ at which the ensemble
is generated. This has been confirmed [\ref{BarbourBell}], [\ref{Tassos}]
on $4^4$ lattices.
However, the rate of convergence of different coefficients
can depend on the updating mass. We choose $m_0a$ to be close
to the expected $m_ca$ observed in other fixed temperature
simulations so
that the coefficients which are the most significant in the
Yang-Lee zeros analysis are obtained with the highest
precision.
\par We analyze the distribution of the zeros of the partition
function in the complex $ma$-plane. We define
the critical value of the bare
mass parameter to be the real part of the zeros closest
to the real axis.
The zeros are obtained from different truncations of the averaged
characteristic polynomial
as the degree
of the full one, ($n_s*nc/2$), is too high for reliable
numerical evaluation of its zeros.
This
procedure requires additional tests of convergence for identification
of numerically accurate zeros. These tests are performed by a procedure
described in [\ref{BarbourBell}] where the partition function
is expanded around shifted values of the mass and its zeros
are reconstructed with respect to these shifts. The zeros
which show stability in expansions with
different shifts are considered to be found with satisfactory
accuracy provided they are also independent of the truncation
procedure ( above some minimum truncation).
The accuracy of the calculation is also checked by
evaluation of the residue of the full polynomial at each zero.
\par We extend the previous evaluation of the critical mass
[\ref{BarbourBell}]
in the intermediate coupling regime on a $4^4$ lattice.
The new
values of the coupling are
$\beta = 4.9$ and $\beta= 5.2$.
The averaged coefficients
of the partition function for $\beta=4.9$
are obtained
from 3260 configurations generated by Hybrid MC at updating mass
$m_0a=0.025$. At $\beta=5.2$ the coefficients are obtained from 1098
configurations with $m_0a=0.18$.
The roots in the vicinity of the update mass do stabilize reasonably
quickly
as the number of configurations contributing
to the average increases.
\par We present in Figs. 1a-1c the distribution of the roots near to
the real axis in the complex $ma$ plane
for $\beta=4.9$, $\beta=5.04$
[\ref{BarbourBell}]
and $\beta=5.2$.
For $\beta =4.9$ and $\beta=5.04$ we see a clear signal of
a possible phase transition around $m_ca=0.01$ and $m_ca=0.04$
respectively. The distribution
at $\beta=5.2$ is completely different in that the nearby roots do not
show any signal of pinching the real axis.
We conclude
that this regime could indicate the smooth crossover
behaviour of the chiral condensate expected
at high temperature
[\ref{GottliebLat}].
Note that this value
of $\beta$ on a $4^4$ lattice is lower than that of the
deconfinement phase transition in quenched QCD
which is at $\beta=5.6$ on a $4^4$ lattice [\ref{Gupta2}].
Therefore, evaluation of the maximal value of
the bare mass for which a possible phase transition can
occur will require
a more detailed study of the distribution between $\beta=5.04$
and $\beta=5.2$.
\bigskip
\newline
\noindent{\bf 3. The distribution of zeros at intermediate coupling
$\beta=5.04$ as a function of lattice volume.}
\bigskip
\newline\noindent
\par In order to examine the finite volume dependence of
our results we have investigated the dependence of the distribution
on the spatial volume
but keeping the temporal length, and hence the
temperature in lattice units, fixed. Results are presented
for $\beta=5.04$ on $6^3\times 4$ and $8^3\times 4$
lattices. The update mass was chosen to be $m_0a=.04$ in the light
of the results at this coupling on a $4^4$ lattice
[\ref{BarbourBell}].
The $6^3\times 4$ results were obtained from 404
configurations and the $8^3\times 4$ results from
262 configurations.
The appropriate choice of the updating mass enables us to extract the
zeros of the partition function, closest to the critical region, from
a smaller sample than the one required in the case described in
[\ref{BarbourBell}]
with $m_0a=0.025$. We find the physically relevant
zeros tending towards the real axis at $m_ca=0.043 \pm 0.001$ on each of
the three lattices.
In Figs.2a-2c we present the distribution in
the complex $ma$-plane of the zeros close to the real axis
for the three lattices. Again the stability of these zeros against
varying the truncation and varying the mass shift in the
characteristic polynomial has been confirmed. Indeed we find
that these nearby zeros are given accurately by approximately the first
${{n_s*n_c/2}\over {10}}$
coefficients of the averaged characteristic polynomial.
It is these coefficients which involve the largest loops on the
lattice and hence reflect, in part, the non-perturbative content
of the theory. We therefore believe that the distribution and density
of these zeros can be studied on much larger lattices using the above
techniques. The other zeros are, in general, sensitive to the
truncation and cannot be obtained with certainty, but their behaviour
under varying the truncation is always consistent with their
distribution being relatively distant from the real axis.
\par In Fig.3 we superimpose the nearby zeros from the three
lattices. They form a curve which appears to be invariant under
increase in spatial volume. It is remarkable that the physics on
these lattices is dominated by a very small fraction of the total
number of zeros. It is also clear from the Fig.3 that $m_ca$ is
stable under increase of the lattice size and that the imaginary part
of the closest zero decreases with increasing lattice size.
\par We first confirm that it is these zeros which control any
apparent phase transition in the order parameter.
We therefore
evaluate the chiral condensate
as a function of the lattice mass with and without the inclusion of
the nearby zeros.
The chiral condensate is derived from
\begin{equation}\label{chiralcon}
<\bar\psi\psi> = 1/Z \partial_{ma}Z(ma)=\partial_{ma} F(ma)
\end{equation}
where $F(ma)$ is the free energy of the system.
\par Since the partition function factorizes as:
\begin{equation}\label{partfact}
Z(ma)=\prod_i(ma-z_i)
\end{equation}
where the $z_i$ are its complex zeros,
the free energy can be presented as
a sum over distinct contributions
each one depending on a different
zero:
\begin{equation}\label{Freedecomp}
F(ma)=\sum_i \log(ma-z_i)
\end{equation}
and hence the part
responsible for the discontinuity
in the order parameter can be isolated. This separation of the free
energy into singular and analytic pieces
is often used in the phase structure
analysis of a condensed matter
systems [\ref{Itzykson}],[\ref{Fisher}],[\ref{Privman}].
Our conjecture is that the nearby
zeros are indeed those leading to a possible phase transition
In Figs.4 and 5 we present a comparison
of the chiral condensate and susceptibility respectively, on a $8^3\times 4$
lattice at $\beta=5.04$,
derived from the full polynomial and with the contribution of the
nearby zeros removed.
Subtraction of their contribution gives a smooth behaviour for
the order parameter and its susceptibilty. The singular content
of the condensate is clearly associated with the nearby zeros.
A similar behaviour is observed on the $4^4$ and $6^3\times 4$ lattices
but, since the pseudosingular behaviour of the free energy is weak on these
spatial volumes due to the small density of the nearby zeros, the separation
into a smooth component and a singular component is not so clear cut.
\par At this value of the coupling, $\beta=5.04$ we also find
that these zeros lie on the arcs of two circles centered at
$ma=\pm m_ca$ of radius $2m_ca$. This is shown in Fig.3 for
$m_ca=0.043$. Further simulations at different couplings are
necessary to investigate the relevance of this observation.
\par In Fig.6 we compare our results to the
simulations performed by several
other groups in which a critical value of the inverse
coupling $\beta$, $\beta_c$,
 is estimated at fixed values of the bare mass
$ma$
[\ref{Gupta1}],
[\ref{Gupta2}],
[\ref{Gupta3}],
[\ref{Fuku}],
[\ref{KogutSinclair}],
[\ref{Sinclair}],
[\ref{Kogut1}],
[\ref{Bitar}].
These simulations show strong sensitivity to the
size of the spatial lattice, especially between $4^4$ and larger
lattices. Presumably this arises from the fact that, on the $4^4$
lattice,
the singular part of the free energy is controlled by essentially
one complex zero with relatively large imaginary part. On larger
lattices, the singular part of the free energy is more clearly
defined. Our results at $\beta=4.9$ and $\beta=5.04$ are in
good agreement with those found by direct measurement of the
condensate and the behaviour of the Polyakov loop (see Fig.6).
\par We have also identified the crossover value of
the bare mass $ma\approx 0.2$
to be at $\beta \leq 5.2$
(which corresponds to the result obtained
by Gupta {\it et al.} [\ref{Gupta2}])
but base this conclusion
on the distribution of zeros on a $4^4$ lattice. It may
well be that finite size effects are again large and further study is
required.
\par Note that the previous estimates of the critical value
of the inverse coupling are based on measurements of the relevant
order parameters, $<\bar\psi\psi>$ and the Polyakov line.
Therefore, the transition studied in these simulations could
be strongly influenced by the finite volume effects discussed
at finite temperature by Gasser and Leutwyller
[\ref{GasserLeutwyller}].
These estimates
encounter difficulties in
distinguishing between finite volume effects and the possible
finite temperature transition.
Performing the simulations at different spatial volumes
can resolve the problem, but it requires study
of different scaling behaviours of the physical observables with
respect to the spatial volume and with respect to
the temperature.
In our
calculations we predict the location of
the phase transition without an
explicit evaluation of the order parameter.
It appears that
this method is less affected by the finite size of the lattice and
that it provides a clearer route towards the thermodynamical limit.
In the $\beta=5.04$ simulation we find no significant
change in the critical mass as the spatial volume increases
from $4^3$ to $8^3$ with our 'critical' point on the $m_ca-\beta$ plane
(Fig. 6) close to that found on the $8^3\times 4$
[\ref{Kogut1}] and $10^3\times 4$ [\ref{Fuku}] lattices.
This observation could also explain why our 'critical masses'
at $\beta=4.9$
and $\beta=5.04$ on the $4^4$ lattice
are consistently below those found at these couplings on $4^4$
lattices by means of the behaviour of the order parameter
[\ref{Gupta1}],[\ref{Gupta3}],[\ref{Bitar}]
but agree with the $8^3\times 4$ results given in [\ref{KogutSinclair}].
Moreover, since our results are insensitive to the simulation
mass, $m_0a$, we can find (if it exists) a critical mass less than the
minimal mass at which dynamical
simulations are possible.
In the strong coupling simulations
[\ref{BarbourBell}],
$m_ca=0.0$
(for $\beta\leq 3.0$)
was found, while a simulation at such mass
would be meaningless.
Our result predicts a chirally broken phase
at any infinitesimal
mass in the thermodynamical limit
for a finite range of coupling
while
the finite volume effects could still mask the phase transition
in a numerical simulation of the condensate
[\ref{JoliceurMorel}].
\bigskip
\newline
\noindent {\bf 4. Conclusions}
\newline\noindent
 Simulations on lattices
at fixed temperature, but with increasing spatial volume,
appear to indicate the possibility of a chirally related phase
transition in QCD at some critical
non-zero lattice mass. On these finite lattices, the signal for
this transition is
superimposed upon that for
a smooth behaviour in the order parameter, $<\bar\psi\psi>$.
Since
the singular part of the free energy in the thermodynamic limit is
given by
\begin{equation}
\int_0  dz \rho(z) \log{[(ma-m_ca)^2-z^2]}
\end{equation}
then the density
of zeros, $\rho(z)$, providing the singular behaviour of the free
energy, must be non-zero in this limit for the transition to remain.
Our results on the three lattice volumes studied are consistent with
with this behaviour but the method used does indicate that the free
energy derived on lattices
with spatial volumes less than $8^3$ has only a weak pseudo-
singular part. More ambitious simulations will be
required to derive the volume dependence of this part of the
distribution, and hence the density of the nearby zeros
in the thermodynamic limit. It is relevant to note that
no evidence was found for zeros
of the partition function on the imaginary axis and close to the
origin. If this feature of their distribution persists as the lattice
volume is increased then the only possible chirally related
transition in the
thermodynamic limit is at non-zero lattice mass.
\bigskip
\newline\noindent{\bf Acknowledgments}
\bigskip\noindent\newline
We thank C.T.H. Davies, S.J. Hands and D.G. Sutherland for illuminating
discussions and constructive comments in the course of this work.
The numerical work was performed on the Cray XMP at RAL.
\newpage
{\bf \noindent References}
\begin{enumerate}
\item\label{KogutMatsuoka}
{J.B. Kogut, H. Matsuoka, M. Stone,
H.W. Wyld, S. Shenker,
J. Shigemitsu and D.K. Sinclair,
{\it Nucl. Phys.} {\bf B225} [FS9] (1983) 93
.}
\item\label{PisarskyWilczek}
{R.D. Pisarsky and F. Wilczek, {\it Phys. Rev.}{\bf D 29} (1984) 338 .}
\item\label{Gupta1}
{R. Gupta, G. Guralnik,
G.W. Kilcup, A. Patel and S.R. Sharpe,
{\it Phys. Rev. Lett.} {\bf 57} (1986) 2621 .}
\item\label{Gupta2}
{R. Gupta, G. Guralnik, G.W. Kilcup, A. Patel and S.R. Sharpe,
{\it Phys. Lett.} {\bf 201B} (1988) 503 .}
\item\label{Gupta3}
{R. Gupta, G.W. Kilcup, and S.R. Sharpe,
{\it Phys. Rev.} {\bf D38} (1988) 1288.}
\item\label{Fuku}
{M. Fukugita and A. Ukawa,
{\it Phys. Rev.} {\bf D38}
(1988) 1971 .}
\item\label{KogutSinclair}
{J.B. Kogut, E.V.E. Kovacs and D.K. Sinclair, {\it Nucl. Phys.}
{\bf B290} [FS20] (1987) 431 .}
\item\label{Sinclair}
{D.K. Sinclair, {\it Nucl. Phys.}{\bf B17} (Proc. Suppl.) (1990) 554 .}
\item\label{Kogut1}
{J.B. Kogut, {\it Nucl. Phys.}{\bf B270}[FS16] (1986) 169 .}
\item\label{Bitar}
{K. Bitar, A.D. Kennedy, R. Horsley, S. Meyer and P. Rossi,
{\it Nucl. Phys.} {\bf B313} (1989) 348 .}
\item\label{JoliceurMorel}
{T. Jolicoeur and A. Morel, {\it Nucl. Phys.} {\bf B262}  (1985) 627 .}
\item\label{GasserLeutwyller}
{J. Gasser and H. Leutwyler, {\it Phys. Lett.} {\bf 188B} (1987) 477 .}
\item\label{BarbourBell}
{I.M. Barbour and A.J. Bell, {\it Nucl. Phys.}{\bf B372} (1992) 385 .}
\item\label{YangLee}
{C.N. Yang and T.D. Lee, {\it Phys. Rev.} {\bf 87} (1952) 404 ;
{T.D. Lee and C.N. Yang, {\it Phys. Rev.} {\bf 87} (1952) 410 .}
\item\label{Itzykson}
{C. Itzykson, R.B. Pearson and J.B. Zuber, {\it Nucl. Phys.}{\bf B220}
[FS8] (1983) 415 .}
\item\label{Tassos}
{I.M. Barbour, A.J. Bell, M. Bernaschi,
G. Salina, A. Vladikas,
{\bf ROME2 91/29}.}
\item\label{Lan}
{I.M. Barbour, N.-E. Behilil. P.E. Gibbs, G. Schierholz and M. Teper,
{\it The recursion method and its applications}, ed. D.G. Pettifor and
D.L. Weaire, {\bf Springer, Berlin,} (1985).}
\item\label{BarbourDagotto}
{I. M. Barbour,
N. Behilil, E. Dagotto, F. Karsch, A. Moreo,
M. Stone and H.W. Wyld,
{\it Nucl. Phys.} {\bf B275} (1986) 296 .}
\item\label{GottliebLat}
{S. Gottlieb, {\it Nucl. Phys,} {\bf B20}(Proc. Suppl.) (1991) 247 .}
\item\label{Fisher}
{V. Privman and M.E. Fisher, {\it Phys. Rev.}{\bf B30} (1984) 322 ;
K. Binder, M. Nauenberg, V. Privman and A.P. Young, {\it ibid.} {\bf 31}
(1985) 1498 .}
\item\label{Privman}
{M.L. Glasser, V. Privman and L.S. Schulman, {\it Phys. Rev.}{\bf B35}
(1987) 1841 .}
\end{enumerate}
\newpage\noindent
{\bf Figure captions}
\newline
Fig. 1-a Distribution of the zeros of the partition function close to
the real axis for $\beta=4.9$ on a $4^4$ lattice.
\newline
Fig. 1-b The same as Fig. 1-a for $\beta=5.04$.
\newline
Fig. 1-c The same as Fig. 1-a for $\beta=5.2$.
\newline
Fig. 2-a Distribution of the zeros of the partition function close to the
real axis for $\beta=5.04$ on a $4^4$ lattice.
\newline
Fig. 2-b The same as Fig. 2-a on a
$6^3\times 4$ lattice.
\newline
Fig. 2-c The same as Fig. 2-a on a
$8^3\times 4$ lattice.
\newline
Fig. 3 Zeros of the partition function close to the real axis
for $\beta=5.04$;
The results on three lattices are superimposed:
$4^4$ --- $\times$, $6^3\times 4$ --- $\star$, $8^3\times 4$ --- $\circ$,
The smallest zeros are located on circles
(dotted arcs) with centres at
$\pm m_ca$.
\newline
Fig. 4 $<\bar\psi\psi>$ as a function of
$ma$ for $\beta=5.04$ on a
$8^3\times 4$ lattice;
the solid line represents
the condensate derived from the partition function and the dashed
line - the smooth behaviour after subtraction of the contribution of
the zeros close to the real axis.
\newline
Fig. 5 The chiral susceptibility as a
function of $ma$ with $\beta$, lattice
size and notation as in Fig. 4.
\newline
Fig. 6 The critical lattice  mass as a
function of $\beta$:
\newline
our results --- dotted white squares;
Refs.\ref{Gupta1},\ref{Gupta2}, on
a $4^4$ lattice --- black diamonds;
Ref.\ref{Gupta3}, on a $4^4$ lattice --- white squares;
Ref.\ref{Gupta3}, on a $6^34$ lattice --- black squares;
Ref.\ref{Fuku}, on a $8^34$ and $10^34$ lattices --- +;
Ref.\ref{KogutSinclair},\ref{Kogut1}, on a $8^34$ lattice --- $\times$;
Ref.\ref{Sinclair}, on a $12^34$
lattice --- triangle;
Ref.\ref{Bitar}, on a $4^4$ lattice ---
black triangle;
Ref.\ref{Bitar}, on a $6^34$ and $8^34$ lattices --- white diamond.
\end{document}